\documentclass[%
 reprint,
 amsmath,amssymb,
 aps,
 prl
]{revtex4-2}
\usepackage{lipsum}  \usepackage{booktabs}
\usepackage{graphicx}
\usepackage{dcolumn}
\usepackage{bm}
\usepackage[dvipsnames]{xcolor}
\usepackage{siunitx}
\usepackage{hyperref}

\begin{document}


\title{A magneto-optical trap of silver and potassium atoms}
\author{Michael Vayninger$^1$}
\author{Angela Xiang$^1$}
\author{Nachiket D. Bhanushali$^1$}
\author{Xiaoyu Chen$^1$}
\author{Mohit Verma$^1$}
\author{Shaozhen Yang$^1$}
\author{Rohan T. Kapur$^1$}
\author{David DeMille$^{1,2}$}
\author{Zoe Z. Yan$^{1,\dagger}$}

\affiliation{$^1$ James Franck Institute and Department of Physics, The University of Chicago, Chicago, IL 60637, USA\\
$^2$ William H. Miller III Department of Physics and Astronomy, Johns Hopkins University, Baltimore, MD 21218 USA}


\date{\today}

\begin{abstract}
We demonstrate a dual magneto-optical trap of $^{109}$Ag and $^{39}$K. 
For silver, a decreasing-field Zeeman slower loads a MOT of $1.5{\times}10^8$ atoms at a temperature of 0.74(5)\,mK, with laser cooling occurring primarily on the $D_2$ line of $4d^{10}5s\; {}^2S_{1/2}\rightarrow 5p\; {}^2P_{3/2}$ at 328\,nm. 
We create a novel Ag ``dark spot MOT," where shelving the atoms in a dark state enhances the captured atom number by a factor of two and the lifetime by a factor of four.
For potassium, we obtain $2{\times}10^8$ trapped atoms, and further cooling on the $D1$ transition via grey molasses results in a cloud of $1.2{\times} 10^8$ atoms at 7(1)\,$\mu$K.  
We observe evidence of photoionization loss of the K MOT in the presence of Ag laser-cooling light, with implications for optimal dual species loading strategies.
Our results on Ag point to simple and general laser cooling strategies for other coinage metals (Au, Cu).
Furthermore, this work lays the foundation for the production of alkali-coinage metal degenerate quantum mixtures and highly polar molecules.
\end{abstract}

\maketitle

The production of trapped, ultracold atoms has revolutionized the field of atomic and molecular physics, enabling quantum simulators and computers~\cite{Bloch2008, gross2017quantum, saffman2016quantum}, precision tests of fundamental physics~\cite{ye2007precision, Safronova2018}, new standards for metrology~\cite{Ludlow2015}, and the field of ultracold quantum chemistry~\cite{zhao2022quantum, karman2024ultracold}.
The workhorse technique for generating ultracold atoms is the magneto-optical trap (MOT)~\cite{schreck2021laser}, and to date, over thirty elements in the periodic table have been magneto-optically trapped. 
Notably, dual species mixtures of different elements are now routinely produced, leading to rich new physics in the form of interspecies-mediated interactions~\cite{baroni2024quantum}, Fermi and Bose polaron physics~\cite{scazza2022repulsive}, and the assembly of polar diatomic molecules from bialkali mixtures~\cite{ni2008high, cornish2024quantum}.

While most dual species mixtures and assembled polar molecules to date have been prepared from Group I (alkali) elements, there is significant motivation to expand mixture experiments beyond bialkalis.
For example, mixtures with a heavy mass imbalance and/or novel magnetic properties could give rise to qualitatively different features such as possible Efimov clusters~\cite{naidon2017efimov} or a pathway toward relatively high-temperature $p$-wave pairing of fermions~\cite{Wu2016}. 
Another motivation to expand ultracold mixtures beyond the alkali elements is to assemble diatomic molecules with a higher electric dipole moment.
Ultracold polar molecules are becoming an increasingly popular platform for precision measurement \cite{Safronova2018, demille2024quantum}, quantum simulation~\cite{langen2024quantum}, and quantum computation~\cite{cornish2024quantum}. 
Interactions between polar molecules are governed by their intrinsic ground state electric dipole moment;
the most polar molecules are ionically-bonded.
Bialkali molecules typically have a bond that is a mixture of ionic and covalent character, resulting in weak-to-moderate electric dipole moments $d$ (\textit{e.g.}~KRb was first realized, with $d$=0.6\;D~\cite{ni2008high}).
Recently, it was proposed that molecules that combine an alkali metal with a coinage metal (copper, silver, gold) would produce some of the highest achievable dipole moments in diatomic molecules~\cite{Smialkowski} due to the coinage metals' large electron affinity. 
Furthermore, the strong intra-molecular electric field produced in alkali-coinage metal molecules could prove useful for precision measurements of parity and time-reversal violating interactions that may reveal beyond-Standard Model physics~\cite{
fleig2021theoretical,sunaga2019merits,klos2022prospects, marc2023candidate, polet2024p, marc2025semi}.
On the quantum simulation and computation front, stronger inter-molecular dipole-dipole couplings that scale as $d^2$ would boost interaction energy scales and reduce the time scales necessary for entanglement generation.
For potassium-silver molecules (KAg), the predicted dipole moment in the ground electronic state is 8.5\,D~\cite{Smialkowski}.
Furthermore, the existence of numerous stable isotopes ($^{39,40,41}$K and $^{107,109}$Ag) allows the formation of both bosonic and fermionic molecules.

\begin{figure*}[t]
\includegraphics[width=\textwidth]{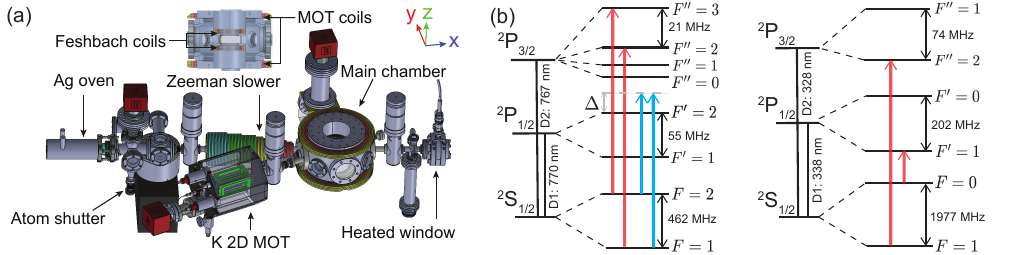}
\caption{\label{fig:1} (a) Illustration of vacuum apparatus, including the Ag effusion oven, Zeeman slower, Ag atom shutter, K 2D MOT, main chamber, magnetic field coils for the MOT, and the heated Ag Zeeman viewport.  The inset shows a cross-section of the main chamber with high numerical aperture viewports and another set of magnetic coils for eventual control over Feshbach resonances. (b) Energy level diagrams for $^{39}$K and $^{109}$Ag, with lasers responsible for the MOT and grey molasses indicated as red and blue arrows, respectively.  Note Ag’s inverted hyperfine structure.
}
\end{figure*}

Ag has a single $5s$ electron over a closed $4d^{10}$ shell in its ground state configuration, enabling optical cycling and laser cooling on a strong transition.
Both stable isotopes of Ag have nuclear spin $I=1/2$, simplifying the hyperfine structure.
Two primary challenges of producing cold Ag are its low vapor pressure at room temperature and its ultraviolet (328\,nm) laser cooling transition.
Despite these difficulties, the Walther group at MPQ~\cite{Uhlenberg2000} created a MOT of $3{\times} 10^6$ atoms by loading out of a thermal source, without any beam slowing.

Here we report on the first realization of dual species cooling and trapping of $^{39}$K and $^{109}$Ag. This represents the first experimental step to achieving degenerate gas mixtures of alkali-coinage metal atoms as well as KAg molecules.
We describe the apparatus, detailing the vacuum, magnetic field, and optical systems for laser cooling and trapping the two species.
Specifically, we introduce two features for Ag -- Zeeman slowing of the atomic beam and a dark spot MOT \cite{Ketterle1993} -- that boost trapped atom numbers to levels commensurate with future creation of quantum-degenerate Ag gases.
Finally, we discuss the results of our optimal cooling strategies for Ag and K.

\begin{figure}[t]
\centering
    \begin{minipage}{0.453\textwidth}
        \centering
        \includegraphics[width=1\textwidth]{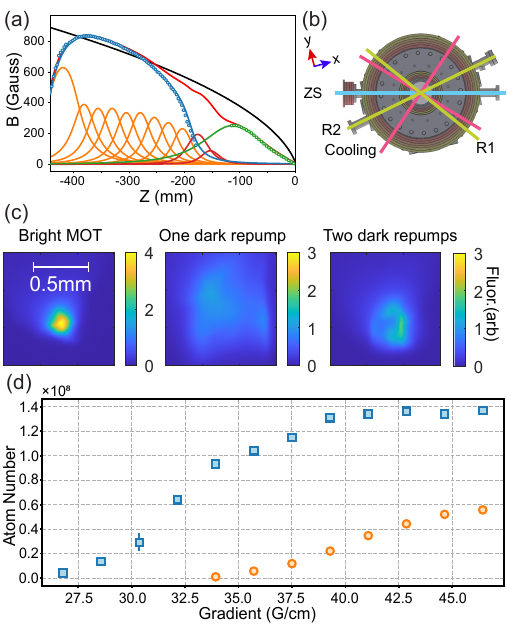} 
    \end{minipage}
\caption{\label{fig:2} Ag slowing and magneto-optical trapping. 
(a)	Magnetic field profile of the Ag Zeeman slower, comprising the simulated profile of 9 segments of the Zeeman solenoid (orange), 2 segments of a ``bridge" solenoid (dark red), and the MOT gradient (green). The combined Zeeman field is shown in blue, and the total field is shown in red, emulating the ideal profile (black).  The measured fields are shown as circles.
(b)	Top-down view of beams used for trapping Ag, including the slowing beam (ZS, blue), MOT cooling (pink), and two repump beams with imaged dark spots (yellow, R1 and R2).
(c)	Fluorescence images of the bright Ag MOT with two unmasked repumping beams, a sub-optimal dark MOT with only one repumper with a dark spot mask, and an optimized dark MOT with two repumpers with intersecting dark spots.  
(d) Ag atom number captured in the bright (orange circles) and dark MOT (blue squares) as a function of axial magnetic field gradient. Error bars are standard errors of the mean.
}
\end{figure}

Our vacuum apparatus contains a two-dimensional MOT and a Zeeman slower that connect the K and Ag sources, respectively, to a common science chamber; see Fig.~\ref{fig:1}(a). 
The K 2D MOT is commercially sourced from SYRTE.
The Ag source is an effusion cell that is heated to 950$^\circ$C (Createc).  The Ag oven region is pumped by two ion pumps, keeping the region below $10^{-9}$ torr.
The two species enter the octagonal science chamber at a relative angle of 26$^\circ$ via a custom 3x1 Conflat flange adapter, thus preserving optical access in the science chamber by only requiring one out of the eight lateral viewports.
The flux of hot Ag atoms is controlled by a homemade atom shutter---a polished, rectangular piece of aluminum---which is mounted on a flange via a bellows and actuated by a solenoid. 
Angled at 45$^\circ$ relative to the Zeeman slowing beam, this shutter doubles as a mirror for coarse alignments of the slowing beam.
In between the atom shutter and the Zeeman slower, we have included four viewports allowing future transverse cooling.
The Zeeman slower has a decreasing-field design, ensuring that the magnetic field merges with the MOT radial field as it drops to zero at the center of the science chamber.
The combined magnetic field profile is shown in Fig.~\ref{fig:2}(a).
The solenoid comprises hollow-core copper magnet wire epoxied together and mounted on a brass tube.
Compared to a spin-flip design, the decreasing-field slower occupies a shorter footprint and requires a Zeeman slowing beam tuned closer to the Ag cycling transition.
Near the main chamber, an additional ion pump ensures ultrahigh vacuum conditions below $10^{-11}$ torr.

Finally, the vacuum system for Ag terminates in a commercial heated window (Thermionics VHW-150).  
Many Zeeman slowers experience deposition of solid material over time on the interior of the vacuum viewport, degrading transmission of the slowing beam.
Unlike alkali metals, whose depositions can be removed by heating the viewport to $\sim 200^\circ$C (a typical limit for vacuum glass-to-metal seals), silver depositions require a much hotter temperature.
Therefore, the heated window features a double-glass design, in which the interior fused silica window lies entirely in vacuum and can reach 700$^\circ$C using in-vacuum heating filaments.  
The outer viewport experiences no significant temperature gradient due to the in-built water-cooling on the flange.
We measure 80\% transmission efficiency of the ultraviolet light through the entire heated window assembly, and the deposition can be periodically removed by heating the assembly to 700$^\circ$C for $\sim$24 hours.
The assembly can be decoupled from the main chamber with a gate valve and pumped with a titanium sublimation pump. 
We expect that this heated viewport could be utilized for other species where Zeeman slowing is known to produce a heavy deposition on vacuum viewports, $\textit{e.g.},$ dysprosium or lithium, as well as for future experiments on other coinage metals.

The laser transitions responsible for the cooling of K and Ag are shown in Fig.~\ref{fig:1}(b).
For K, cooling and trapping rely on the $D_2$ transition at 766.7\,nm with natural linewidth $\Gamma_{\rm K}/2\pi=6.035$\,MHz~\cite{PhysRevA.74.032503}.
A fiber laser (Shanghai Precilaser) of 2.5\,W generates the cooling and repumping beams, which are detuned from their respective transitions by -8.5 and -8 $\Gamma_{\rm K}$.
With various acousto-optic modulators, we generate light and fiber couple 160(190) mW of cooling (repump) to the K 2D MOT, 180(70) mW of cooling (repump) to the 3D MOT, and 0.4 mW of resonant light to ``push” the atoms from the 2D MOT chamber to the main chamber. 
For the 3D MOT, the cooling and repumping light are independently fiber-coupled, and then the outputs are combined on a 50:50 non-polarizing beam splitter to generate the MOT beams (which are further split by polarizing beam splitters to generate three beams). 
The 3D MOT beams each have a peak intensity $I{=} 84\,I_{\rm 0,K}$, where $I_{\rm 0,K}$=1.75\,mW/cm$^2$ is the saturation intensity of potassium. 
The MOT beams are each retroreflected to form six total beams.
Laser stabilization is achieved for the $D_2$ line via Doppler-free modulation transfer spectroscopy on a vapor cell.

Additionally, sub-Doppler cooling is performed on the $D_1$ line at 770.1\,nm, with up to 1.5\,W of power generated by a diode laser seeding a tapered amplifier (Toptica Eagleyard ECL, Mini-TA). 
This light is combined on an interference filter with the 3D repumper before fiber coupling, ensuring that the $D_1$ light is spatially superimposed with the $D_2$ MOT trapping light.
We implement $\Lambda$-enhanced grey molasses~\cite{nath2013, salomon2014gray}, operating with a single-photon detuning of $\Delta{=} 5\,\Gamma_{\rm K}$, and intensities of 3.5\,$I_{\rm 0,K}$ and 1.2\,$I_{\rm 0,K}$ for the cooling and repumping transitions, respectively.
For the $D_1$ laser stabilization, we reference the seed laser to a wavemeter (HighFinesse WS7), which is stored in a vacuum canister for thermal and barometric stability.

Ag’s electronic configuration is highly comparable to an alkali metal, with $D_2$ and $D_1$ transitions from the $S_{1/2}{\rightarrow} P_{3/2},P_{1/2}$ at 328\,nm and 338\,nm, respectively~\cite{badr2004continuous} (see Fig.~\ref{fig:1}(b)). 
To generate 328\,nm, two fiber lasers combine to form over 4\,W of 656\,nm through sum-frequency generation (Shanghai Precilasers). 
Then, this light is frequency-doubled to the ultraviolet via second harmonic generation (Agile Optic), producing maximally 1.2\,W of 328\,nm, but with typical daily operation of $\sim$600-900\,mW.
The $D_1$ light is produced with a second, analogous system.
The short wavelength of the $D_2$ and $D_1$ transitions result in high natural linewidths $\Gamma_{\rm Ag~ D2 (D1)} {=} \,23.4 (21.5)$\,MHz and saturation intensities $I_{\rm 0,Ag~ D2 (D1)} {=}\, 87 (73)$\,mW/cm$^2$.
Cooling and slowing are performed with respect to the $D_2~F{=}1{\rightarrow} F''{=}2$ cycling transition, with frequency shifting and amplitude modulation accomplished with AOMs.
Due to the spontaneous decay of off-resonantly-excited $F''{=}1$ to the ground $F{=}0$ dark state, we employ a $D_1$ repumper in the $F{=}0{\rightarrow} F’{=}1$ transition. 
Both the $D_2$ and $D_1$ lasers are frequency-stabilized on the same wavemeter as the K $D_1$ laser.
To rapidly switch between the three sources, we built a custom fiber switch using a galvanometer (Thorlabs GVSK1) that alternates coupling the 770\,nm, 656\,nm, and 676\,nm beams into a single fiber into the wavemeter. 
A digital PID feedback controller ensures that the root-mean-square frequency deviations are 1.5, 0.5, and 0.5 MHz for these three lasers, respectively.
The wavemeter is periodically calibrated to our 767\,nm K $D_2$ laser that is stabilized to atomic vapor, ensuring an absolute frequency calibration at the $\sim$MHz level.

The 3D MOT beams of Ag are generated by splitting the MOT light with polarizing beam splitters, and are combined with the K MOT beams via long-pass dichroic filters (Edmund 86-330). All subsequent mirrors and vacuum viewports have dual AR-coatings at 328\,nm and 767\,nm.
Like the K 3D MOT beams, the Ag MOT beams are all retroreflected. 
The MOT beams are generated by reflecting the incident beam through a $\lambda/4$ waveplate (to switch polarization), a positive focal-length $f$ lens, and a mirror placed one focal length away from the lens $\sim f$.
The lens position allows fine tuning of the retroreflected beam size and therefore intensity, which could otherwise be attenuated by transmission loss through the chamber viewports.
We separate the K and Ag beams after exiting the chamber via another dichroic filter, and retroreflect with independent lenses, allowing for quasi-independent tuning of all twelve MOT beam intensities for Ag and K.

Here we describe slowing and magneto-optical trapping of Ag, including a ``dark spot MOT”~\cite{Ketterle1993} that more than doubles the atom numbers.
First, our conventional ``bright” MOT of atoms is achieved by loading atoms with 55 mW of slowing light detuned at -1.6 $\Gamma_{\rm Ag,D2}$.
The beam waist is 7\,mm at the MOT and focuses down to 2\,mm at the exit aperture of the effusion cell.
Due to the Zeeman structure of the $F=0,1$ states, we are unable to simultaneously design the decreasing-field electromagnet to fulfill the Zeeman and Doppler resonant conditions for both the cooling and repumping transitions, for a particular velocity of atoms.
Therefore, we forego repumping in the slower and expect that we capture fewer than $1/4$ of the initial $^{109}$Ag population due to thermal Boltzmann distribution among the hyperfine states and subsequent leakage out of the cycling transition.
Further improvements to the loading rate could be made by adding a dedicated repumping beam to the slower.

\begin{figure}[t]
    \centering
    \begin{minipage}{0.453\textwidth}
        \centering
        \includegraphics[width=1\textwidth]{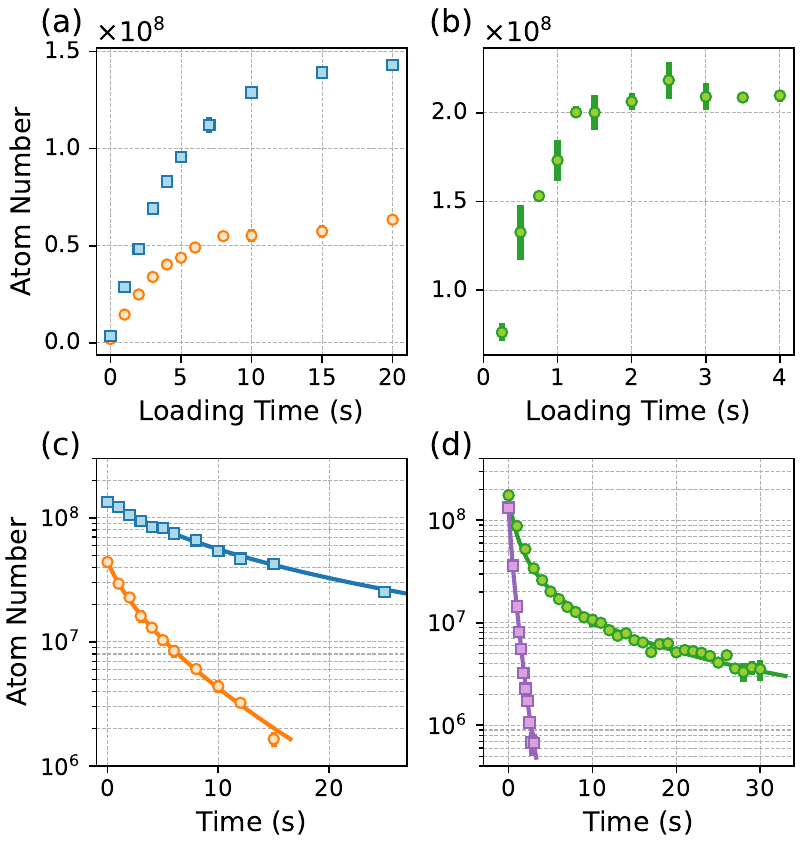} 
    \end{minipage}
\caption{\label{fig:3} MOT loading rate and lifetimes, showing atom number during loading for (a) Ag bright and dark MOTs in orange circles and blue squares, respectively and (b) K MOT. 
(c) Lifetime of the Ag MOT measured after turning off the slow beam via the atom shutter, with fitted one-body loss rates of $\alpha{=}1.5(4) {\times}10^{-1}$\,/s and $3.6(5) {\times}10^{-2}$\,/s and two-body loss rates of $\beta{=}7(1){\times}10^{-11} \: \rm{cm^3/s}$ and $3.5(6) {\times} 10^{-13}\: \rm{cm^3/s}$ for the bright and dark MOTs, respectively (see text). 
(d) Lifetime
of K after turning off the 2D MOT source, both with (green) and without (purple) UV light, leading to one-body rates of $\alpha{=}1.8(1)$\,/s and $\alpha{=}4.0(4)\times10^{-2}$\,/s, respectively. }
\end{figure}

The Ag 3D MOT is created with 6 cooling beams detuned at -0.9 $\Gamma_{\rm Ag,D2}$, with average single beam intensity of 0.9\,$I_{\rm 0,Ag,D2}$.
We find the greatest atom number at a magnetic field gradient of $\frac{dB}{dz} {=} 46$\,G/cm (see Fig.~\ref{fig:3}(d)), over 4 times higher than the field preferred by our K MOT.
Repumping is achieved via two beams of the same frequency on two different axes (see Fig.~\ref{fig:2}(b-c)), with detunings of -3 $\Gamma_{\rm Ag, D1}$, and intensities of 0.2$\,I_{\rm 0,Ag,D1}$.
Three pairs of bias fields in $x, y, z$ ensure that the magnetic gradient is centered where the beams overlap. 
For the bright MOT, we typically load $0.6{\times}10^8$ atoms in $\sim 10$\,s, as pictured in Fig.~\ref{fig:3}(a). 
The final number is likely limited by decay from the excited $P_{3/2}$ to a metastable $D_{5/2}$ state with a lifetime of 5\,s~\cite{Dzuba2021}, long enough for the atoms to fall out of the trap.
Instead of adding additional lasers to repump this state, we pursue a different strategy to mitigate these losses: a dark spot MOT.

The dark spot MOT was first demonstrated successfully on a sodium MOT~\cite{Ketterle1993}, where masking the central portion of a repumping beam allowed the center of the MOT to fall into a dark state. 
With radiation forces almost absent, there are far fewer scattering and reabsorption events, allowing a strong density enhancement compared to a bright MOT.
In the case of Ag, not only would the dark spot eliminate most of the detrimental rescattering events, but it would additionally mitigate the decay to the $D_{5/2}$ state. 
We implement dark spots imaged onto the center of each $D_1$ repumping beam, with R1 (R2) having central spot diameters of 1.5 (1.6) mm at the MOT center, respectively. 
The masks are generated from black circular stickers and mounted on a $xy$ translation stage to overlap with the bright MOT.
With the dark MOT, our atom number improves to 1.5${\times }10^8$ loaded in 20\,s (see Fig.~\ref{fig:3}(a)).
We also measure the lifetime of the MOTs and fit decay curves to the phenomenological loss function $\frac{dN}{dt} {=} {-}\alpha N {-} \beta Nn$, where $N$ is the number, $n$ is the average number density, $\alpha$ is the one-body loss rate, and $\beta$ is the two-body loss rate. The dark MOT improves the one-body lifetimes by a factor of four compared to the bright MOT (see Fig.~\ref{fig:3}(c)). 
The dark MOT's dramatic improvement in both the one- and two-body loss rates suggests that the dark spot mitigates population leakage out of the cooling states and suppresses inelastic light-assisted collisions.

\begin{figure}[t]
\centering
    \begin{minipage}{0.453\textwidth}
        \centering
        \includegraphics[width=\textwidth]{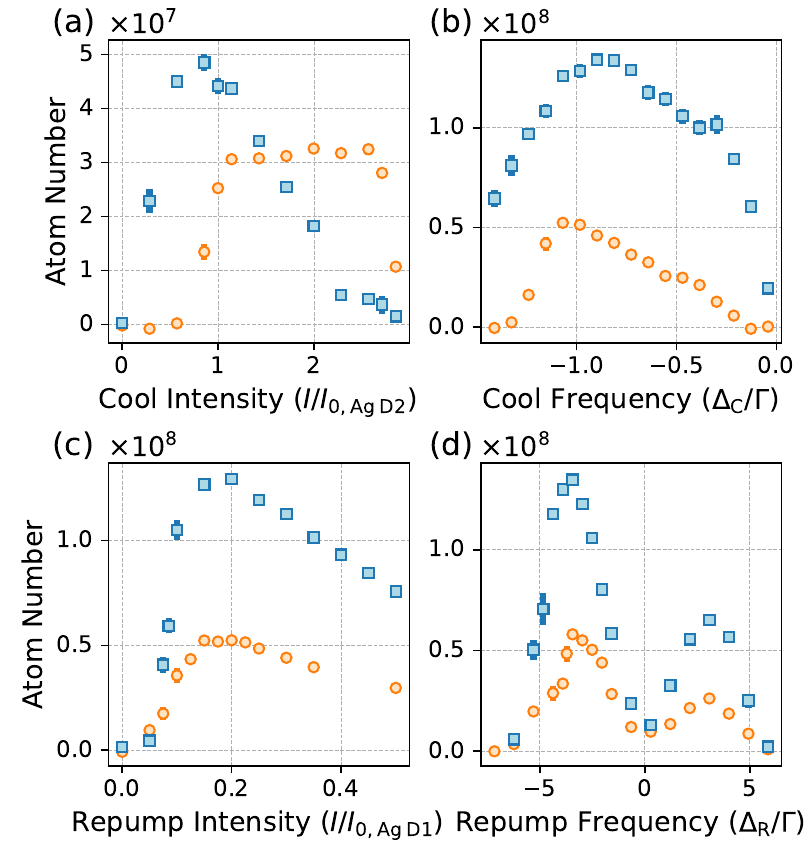}
    \end{minipage}
\caption{\label{fig:4} Steady-state Ag population for the bright MOT (orange circles) and dark MOT (blue squares) as a function of (a) cooling intensity, (b) detuning of the cooling beam $\Delta_{\rm C}$, (c) repump intensity, and (d) repump detuning $\Delta_{\rm R}$.
As our limited UV power is split between our Zeeman slower and cooling beams, we scan the cooling intensity in (a) with the Zeeman beam at 1/5 of the intensity that optimizes the MOT. 
For (b-d), the non-scanned parameters were held constant and set to optimize atom number.
The MOT gradient was set to 46\,G/cm for all measurements.}
\end{figure}

Furthermore, we study the effect of cooling and repump intensities and detunings on the dark MOT atom number, as shown in Fig.~\ref{fig:4}, after 10\,s of loading. 
The optimal cooling detuning lies only -0.9\,$\Gamma_{\rm Ag,D2}$ away from resonance, a small value compared to typical alkali MOTs, due to our limited ultraviolet power and operating at only $I/I_{\rm 0, Ag,D2} {=}0.9$ with 1 cm beam diameters. 
For the repumping beams, laser power is not a limit and a two-dimensional optimization of intensity and detuning produces the best results at $I/I_{\rm 0, Ag,D1}{ =} 0.2$ and -3\,$\Gamma_{\rm Ag,D1}$.
The coldest temperatures reached are 740(50)\,$\mu$K compared to the Doppler temperature of $\hbar\Gamma_{\rm Ag, D2}/2=561\,\mu$K (see Tab.~\ref{tab}). 
Our optimal densities for the dark MOT are $8(1){\times} 10^{11}$\,/cm$^{3}$.

\begin{table}[h]
\vspace{0.5em}
\centering
\begin{tabular}{lcccc}
\toprule
 & K & K Molasses & Ag Bright & Ag Dark \\
\midrule
Number ($10^8$)       & 2.0 & 1.2 & 0.6 & 1.5 \\
Temperature (mK)       & 9.1(8) & 0.007(1) & 2.3(3) & 0.74(5) \\
\bottomrule
\end{tabular}
\caption{\label{tab} Numbers and temperatures achieved for the K 3D MOT and D1 molasses, Ag in the bright MOT, and Ag in the dark spot MOT.}
\end{table}

In a separate but similar apparatus, we have also trapped Ag under conditions optimized for single-species trapping. There, the dark spot MOT (formed by repumping with the D1 laser or sidebands of the D2 laser) contains $9.6{\times}10^8$ Ag atoms, at density $2.0(2){\times}10^{11}$\,/cm$^{3}$ and temperature 2.3(2)\,mK, with details in a forthcoming manuscript.

The cooling and trapping of K is more well-established than that of Ag, so we describe our results in brief.
The 3D MOT saturates in 2.5\,s to $>2{\times}10^8$ atoms, as shown in Fig.~\ref{fig:3}(b). 
Average MOT temperatures are 9.1(8)\,mK compared to the Doppler temperature of 145 $\mu$K.
The optimal magnetic gradient for loading is 11 G/cm, with higher gradients producing fewer atoms at hotter temperatures.
MOT densities are enhanced with compression, whereby we turn the cooling frequency to -4.5$\Gamma_{\rm K}$, and the repump intensity to 1.4 $I_{\rm 0,K}$.
This results in suppressed temperatures of 250(40)\,$\mu$K and enhanced densities. 
A final stage of $D_1$ grey molasses~\cite{nath2013, salomon2014gray} brings the temperature to 7(1) $\mu$K, with the final atom number of 1.2${\times}10^8$ and density of $4.8(5)\times 10^{10}$/cm$^3$.

We observe a strong influence of the Ag laser cooling light on the K MOT, likely due to the 328\,nm light lying beyond the ionization threshold of the K $P_{3/2}$ atoms. 
At the optimal K MOT parameters, the lifetimes of the K MOT with and without the ultraviolet light (at the preferred intensity for the Ag MOT) are shown in Fig.~\ref{fig:3}(d).
In the presence of ultraviolet light, the K loss is strongly one-body-loss dominated with $\alpha {=} 1.8(1)$\,/s, and $\beta$ is not resolved within our signal-to-noise.
Without ultraviolet light, the one-body loss rate falls to $\alpha {=}4.0(4){\times}10^{-2}$\,/s, likely attributable to background gas collisions.
The negative effect of ultraviolet light necessitates sequential loading, with Ag loaded first.
We note that there is no apparent effect of the K laser cooling light on the Ag MOT.

In conclusion, we have presented an apparatus capable of cooling and trapping K and Ag atoms, incorporating a novel Zeeman slower and dark spot repumping design for trapping of 150 million Ag atoms. 
While simultaneous MOT loading of the two species is not efficient (similar to many two-species mixtures), we expect that first loading and optically trapping Ag, followed by MOT loading of K and independent optical trapping, will lead toward dual species quantum degeneracy. 
Furthermore, sub-Doppler cooling of Ag with $D_1$ $\Lambda$-enhanced grey molasses is expected to lower cloud temperatures.
We note that Ag’s scalar ac polarizability at 1064\,nm is 53.8 a.u.~\cite{Porsev2025}, which is quite weak compared to the alkalis, but a high power infrared laser is expected to generate intensities sufficient for optical trapping. 
Our temperatures, atom numbers, and densities provide initial conditions favorable for subsequent evaporative cooling toward the realization of degenerate alkali-coinage metal mixtures and the production of highly polar molecules.

We would like to thank Jack Dewhurst, Tanvi Rao, Ken Liu, Ana Elias, and Zhuoli Ding for their experimental assistance. We thank Thomas Langin, Alan O. Jamison, Micha\l~Tomza, and Mariusz Semczuk for helpful discussions. The experimental work was supported by the Neubauer Family Assistant Professors Program, the David and Lucile Packard Foundation (grant 2024-77404) and NSF grant PHY-2208024. 
Mo.V. acknowledges support from NSERC.

Mi.V., A.X., N.B., X.C., Z.Y., Mo.V., S.Y., and R.K.~built the experimental apparatus. Mi.V., A.X., Z.Y., Mo.V., and S.Y.~performed the experiments and analyzed the data.
Z.Y.~conceived the study. Z.Y.~and D.D.~supervised the experiment. All authors contributed to writing the manuscript.

\noindent $^\dagger$ Email: zzyan@uchicago.edu

\bibliography{bib}

\begin{thebibliography}{32}%
\makeatletter
\providecommand \@ifxundefined [1]{%
 \@ifx{#1\undefined}
}%
\providecommand \@ifnum [1]{%
 \ifnum #1\expandafter \@firstoftwo
 \else \expandafter \@secondoftwo
 \fi
}%
\providecommand \@ifx [1]{%
 \ifx #1\expandafter \@firstoftwo
 \else \expandafter \@secondoftwo
 \fi
}%
\providecommand \natexlab [1]{#1}%
\providecommand \enquote  [1]{``#1''}%
\providecommand \bibnamefont  [1]{#1}%
\providecommand \bibfnamefont [1]{#1}%
\providecommand \citenamefont [1]{#1}%
\providecommand \href@noop [0]{\@secondoftwo}%
\providecommand \href [0]{\begingroup \@sanitize@url \@href}%
\providecommand \@href[1]{\@@startlink{#1}\@@href}%
\providecommand \@@href[1]{\endgroup#1\@@endlink}%
\providecommand \@sanitize@url [0]{\catcode `\\12\catcode `\$12\catcode `\&12\catcode `\#12\catcode `\^12\catcode `\_12\catcode `\%12\relax}%
\providecommand \@@startlink[1]{}%
\providecommand \@@endlink[0]{}%
\providecommand \url  [0]{\begingroup\@sanitize@url \@url }%
\providecommand \@url [1]{\endgroup\@href {#1}{\urlprefix }}%
\providecommand \urlprefix  [0]{URL }%
\providecommand \Eprint [0]{\href }%
\providecommand \doibase [0]{https://doi.org/}%
\providecommand \selectlanguage [0]{\@gobble}%
\providecommand \bibinfo  [0]{\@secondoftwo}%
\providecommand \bibfield  [0]{\@secondoftwo}%
\providecommand \translation [1]{[#1]}%
\providecommand \BibitemOpen [0]{}%
\providecommand \bibitemStop [0]{}%
\providecommand \bibitemNoStop [0]{.\EOS\space}%
\providecommand \EOS [0]{\spacefactor3000\relax}%
\providecommand \BibitemShut  [1]{\csname bibitem#1\endcsname}%
\let\auto@bib@innerbib\@empty
\bibitem [{\citenamefont {Bloch}\ \emph {et~al.}(2008)\citenamefont {Bloch}, \citenamefont {Dalibard},\ and\ \citenamefont {Zwerger}}]{Bloch2008}%
  \BibitemOpen
  \bibfield  {author} {\bibinfo {author} {\bibfnamefont {I.}~\bibnamefont {Bloch}}, \bibinfo {author} {\bibfnamefont {J.}~\bibnamefont {Dalibard}},\ and\ \bibinfo {author} {\bibfnamefont {W.}~\bibnamefont {Zwerger}},\ }\href {https://doi.org/10.1103/RevModPhys.80.885} {\bibfield  {journal} {\bibinfo  {journal} {Rev. Mod. Phys.}\ }\textbf {\bibinfo {volume} {80}},\ \bibinfo {pages} {885} (\bibinfo {year} {2008})}\BibitemShut {NoStop}%
\bibitem [{\citenamefont {Gross}\ and\ \citenamefont {Bloch}(2017)}]{gross2017quantum}%
  \BibitemOpen
  \bibfield  {author} {\bibinfo {author} {\bibfnamefont {C.}~\bibnamefont {Gross}}\ and\ \bibinfo {author} {\bibfnamefont {I.}~\bibnamefont {Bloch}},\ }\href {https://www.science.org/doi/10.1126/science.aal3837} {\bibfield  {journal} {\bibinfo  {journal} {Science}\ }\textbf {\bibinfo {volume} {357}},\ \bibinfo {pages} {995} (\bibinfo {year} {2017})}\BibitemShut {NoStop}%
\bibitem [{\citenamefont {Saffman}(2016)}]{saffman2016quantum}%
  \BibitemOpen
  \bibfield  {author} {\bibinfo {author} {\bibfnamefont {M.}~\bibnamefont {Saffman}},\ }\href {https://iopscience.iop.org/article/10.1088/0953-4075/49/20/202001/meta} {\bibfield  {journal} {\bibinfo  {journal} {Journal of Physics B: Atomic, Molecular and Optical Physics}\ }\textbf {\bibinfo {volume} {49}},\ \bibinfo {pages} {202001} (\bibinfo {year} {2016})}\BibitemShut {NoStop}%
\bibitem [{\citenamefont {Ye}\ \emph {et~al.}(2007)\citenamefont {Ye}, \citenamefont {Blatt}, \citenamefont {Boyd}, \citenamefont {Foreman}, \citenamefont {Hudson}, \citenamefont {Ido}, \citenamefont {Lev}, \citenamefont {Ludlow}, \citenamefont {Sawyer}, \citenamefont {Stuhl} \emph {et~al.}}]{ye2007precision}%
  \BibitemOpen
  \bibfield  {author} {\bibinfo {author} {\bibfnamefont {J.}~\bibnamefont {Ye}}, \bibinfo {author} {\bibfnamefont {S.}~\bibnamefont {Blatt}}, \bibinfo {author} {\bibfnamefont {M.~M.}\ \bibnamefont {Boyd}}, \bibinfo {author} {\bibfnamefont {S.~M.}\ \bibnamefont {Foreman}}, \bibinfo {author} {\bibfnamefont {E.~R.}\ \bibnamefont {Hudson}}, \bibinfo {author} {\bibfnamefont {T.}~\bibnamefont {Ido}}, \bibinfo {author} {\bibfnamefont {B.}~\bibnamefont {Lev}}, \bibinfo {author} {\bibfnamefont {A.~D.}\ \bibnamefont {Ludlow}}, \bibinfo {author} {\bibfnamefont {B.~C.}\ \bibnamefont {Sawyer}}, \bibinfo {author} {\bibfnamefont {B.}~\bibnamefont {Stuhl}}, \emph {et~al.},\ }\href {https://pubs.aip.org/aip/acp/article/869/1/80/1009732/Precision-Measurement-Based-on-Ultracold-Atoms-and} {\bibfield  {journal} {\bibinfo  {journal} {International Journal of Modern Physics D}\ }\textbf {\bibinfo {volume} {16}},\ \bibinfo {pages} {2481} (\bibinfo {year} {2007})}\BibitemShut {NoStop}%
\bibitem [{\citenamefont {Safronova}\ \emph {et~al.}(2018)\citenamefont {Safronova}, \citenamefont {Budker}, \citenamefont {DeMille}, \citenamefont {Kimball}, \citenamefont {Derevianko},\ and\ \citenamefont {Clark}}]{Safronova2018}%
  \BibitemOpen
  \bibfield  {author} {\bibinfo {author} {\bibfnamefont {M.~S.}\ \bibnamefont {Safronova}}, \bibinfo {author} {\bibfnamefont {D.}~\bibnamefont {Budker}}, \bibinfo {author} {\bibfnamefont {D.}~\bibnamefont {DeMille}}, \bibinfo {author} {\bibfnamefont {D.~F.~J.}\ \bibnamefont {Kimball}}, \bibinfo {author} {\bibfnamefont {A.}~\bibnamefont {Derevianko}},\ and\ \bibinfo {author} {\bibfnamefont {C.~W.}\ \bibnamefont {Clark}},\ }\href {https://doi.org/10.1103/RevModPhys.90.025008} {\bibfield  {journal} {\bibinfo  {journal} {Rev. Mod. Phys.}\ }\textbf {\bibinfo {volume} {90}},\ \bibinfo {pages} {025008} (\bibinfo {year} {2018})}\BibitemShut {NoStop}%
\bibitem [{\citenamefont {Ludlow}\ \emph {et~al.}(2015)\citenamefont {Ludlow}, \citenamefont {Boyd}, \citenamefont {Ye}, \citenamefont {Peik},\ and\ \citenamefont {Schmidt}}]{Ludlow2015}%
  \BibitemOpen
  \bibfield  {author} {\bibinfo {author} {\bibfnamefont {A.~D.}\ \bibnamefont {Ludlow}}, \bibinfo {author} {\bibfnamefont {M.~M.}\ \bibnamefont {Boyd}}, \bibinfo {author} {\bibfnamefont {J.}~\bibnamefont {Ye}}, \bibinfo {author} {\bibfnamefont {E.}~\bibnamefont {Peik}},\ and\ \bibinfo {author} {\bibfnamefont {P.~O.}\ \bibnamefont {Schmidt}},\ }\href {https://doi.org/10.1103/RevModPhys.87.637} {\bibfield  {journal} {\bibinfo  {journal} {Rev. Mod. Phys.}\ }\textbf {\bibinfo {volume} {87}},\ \bibinfo {pages} {637} (\bibinfo {year} {2015})}\BibitemShut {NoStop}%
\bibitem [{\citenamefont {Zhao}\ and\ \citenamefont {Pan}(2022)}]{zhao2022quantum}%
  \BibitemOpen
  \bibfield  {author} {\bibinfo {author} {\bibfnamefont {B.}~\bibnamefont {Zhao}}\ and\ \bibinfo {author} {\bibfnamefont {J.-W.}\ \bibnamefont {Pan}},\ }\href {https://pubs.rsc.org/en/content/articlelanding/2022/cs/d1cs01040a} {\bibfield  {journal} {\bibinfo  {journal} {Chemical Society Reviews}\ }\textbf {\bibinfo {volume} {51}},\ \bibinfo {pages} {1685} (\bibinfo {year} {2022})}\BibitemShut {NoStop}%
\bibitem [{\citenamefont {Karman}\ \emph {et~al.}(2024)\citenamefont {Karman}, \citenamefont {Tomza},\ and\ \citenamefont {P{\'e}rez-R{\'\i}os}}]{karman2024ultracold}%
  \BibitemOpen
  \bibfield  {author} {\bibinfo {author} {\bibfnamefont {T.}~\bibnamefont {Karman}}, \bibinfo {author} {\bibfnamefont {M.}~\bibnamefont {Tomza}},\ and\ \bibinfo {author} {\bibfnamefont {J.}~\bibnamefont {P{\'e}rez-R{\'\i}os}},\ }\href {https://doi.org/10.1038/s41567-024-02467-3} {\bibfield  {journal} {\bibinfo  {journal} {Nature Physics}\ }\textbf {\bibinfo {volume} {20}},\ \bibinfo {pages} {722} (\bibinfo {year} {2024})}\BibitemShut {NoStop}%
\bibitem [{\citenamefont {Schreck}\ and\ \citenamefont {Druten}(2021)}]{schreck2021laser}%
  \BibitemOpen
  \bibfield  {author} {\bibinfo {author} {\bibfnamefont {F.}~\bibnamefont {Schreck}}\ and\ \bibinfo {author} {\bibfnamefont {K.~v.}\ \bibnamefont {Druten}},\ }\href {https://doi.org/10.1038/s41567-021-01379-w} {\bibfield  {journal} {\bibinfo  {journal} {Nature Physics}\ }\textbf {\bibinfo {volume} {17}},\ \bibinfo {pages} {1296} (\bibinfo {year} {2021})}\BibitemShut {NoStop}%
\bibitem [{\citenamefont {Baroni}\ \emph {et~al.}(2024)\citenamefont {Baroni}, \citenamefont {Lamporesi},\ and\ \citenamefont {Zaccanti}}]{baroni2024quantum}%
  \BibitemOpen
  \bibfield  {author} {\bibinfo {author} {\bibfnamefont {C.}~\bibnamefont {Baroni}}, \bibinfo {author} {\bibfnamefont {G.}~\bibnamefont {Lamporesi}},\ and\ \bibinfo {author} {\bibfnamefont {M.}~\bibnamefont {Zaccanti}},\ }\href {https://doi.org/10.1038/s42254-024-00773-6} {\bibfield  {journal} {\bibinfo  {journal} {Nature Reviews Physics}\ }\textbf {\bibinfo {volume} {6}},\ \bibinfo {pages} {736–752} (\bibinfo {year} {2024})}\BibitemShut {NoStop}%
\bibitem [{\citenamefont {Scazza}\ \emph {et~al.}(2022)\citenamefont {Scazza}, \citenamefont {Zaccanti}, \citenamefont {Massignan}, \citenamefont {Parish},\ and\ \citenamefont {Levinsen}}]{scazza2022repulsive}%
  \BibitemOpen
  \bibfield  {author} {\bibinfo {author} {\bibfnamefont {F.}~\bibnamefont {Scazza}}, \bibinfo {author} {\bibfnamefont {M.}~\bibnamefont {Zaccanti}}, \bibinfo {author} {\bibfnamefont {P.}~\bibnamefont {Massignan}}, \bibinfo {author} {\bibfnamefont {M.~M.}\ \bibnamefont {Parish}},\ and\ \bibinfo {author} {\bibfnamefont {J.}~\bibnamefont {Levinsen}},\ }\href {https://doi.org/10.3390/atoms10020055} {\bibfield  {journal} {\bibinfo  {journal} {Atoms}\ }\textbf {\bibinfo {volume} {10}},\ \bibinfo {pages} {55} (\bibinfo {year} {2022})}\BibitemShut {NoStop}%
\bibitem [{\citenamefont {Ni}\ \emph {et~al.}(2008)\citenamefont {Ni}, \citenamefont {Ospelkaus}, \citenamefont {De~Miranda}, \citenamefont {Pe'Er}, \citenamefont {Neyenhuis}, \citenamefont {Zirbel}, \citenamefont {Kotochigova}, \citenamefont {Julienne}, \citenamefont {Jin},\ and\ \citenamefont {Ye}}]{ni2008high}%
  \BibitemOpen
  \bibfield  {author} {\bibinfo {author} {\bibfnamefont {K.-K.}\ \bibnamefont {Ni}}, \bibinfo {author} {\bibfnamefont {S.}~\bibnamefont {Ospelkaus}}, \bibinfo {author} {\bibfnamefont {M.}~\bibnamefont {De~Miranda}}, \bibinfo {author} {\bibfnamefont {A.}~\bibnamefont {Pe'Er}}, \bibinfo {author} {\bibfnamefont {B.}~\bibnamefont {Neyenhuis}}, \bibinfo {author} {\bibfnamefont {J.}~\bibnamefont {Zirbel}}, \bibinfo {author} {\bibfnamefont {S.}~\bibnamefont {Kotochigova}}, \bibinfo {author} {\bibfnamefont {P.}~\bibnamefont {Julienne}}, \bibinfo {author} {\bibfnamefont {D.}~\bibnamefont {Jin}},\ and\ \bibinfo {author} {\bibfnamefont {J.}~\bibnamefont {Ye}},\ }\href {https://doi.org/10.1126/science.1163861} {\bibfield  {journal} {\bibinfo  {journal} {Science}\ }\textbf {\bibinfo {volume} {322}},\ \bibinfo {pages} {231} (\bibinfo {year} {2008})}\BibitemShut {NoStop}%
\bibitem [{\citenamefont {Cornish}\ \emph {et~al.}(2024)\citenamefont {Cornish}, \citenamefont {Tarbutt},\ and\ \citenamefont {Hazzard}}]{cornish2024quantum}%
  \BibitemOpen
  \bibfield  {author} {\bibinfo {author} {\bibfnamefont {S.~L.}\ \bibnamefont {Cornish}}, \bibinfo {author} {\bibfnamefont {M.~R.}\ \bibnamefont {Tarbutt}},\ and\ \bibinfo {author} {\bibfnamefont {K.~R.}\ \bibnamefont {Hazzard}},\ }\href {https://doi.org/10.1038/s41567-024-02453-9} {\bibfield  {journal} {\bibinfo  {journal} {Nature Physics}\ }\textbf {\bibinfo {volume} {20}},\ \bibinfo {pages} {730} (\bibinfo {year} {2024})}\BibitemShut {NoStop}%
\bibitem [{\citenamefont {Naidon}\ and\ \citenamefont {Endo}(2017)}]{naidon2017efimov}%
  \BibitemOpen
  \bibfield  {author} {\bibinfo {author} {\bibfnamefont {P.}~\bibnamefont {Naidon}}\ and\ \bibinfo {author} {\bibfnamefont {S.}~\bibnamefont {Endo}},\ }\href {https://iopscience.iop.org/article/10.1088/1361-6633/aa50e8} {\bibfield  {journal} {\bibinfo  {journal} {Reports on Progress in Physics}\ }\textbf {\bibinfo {volume} {80}},\ \bibinfo {pages} {056001} (\bibinfo {year} {2017})}\BibitemShut {NoStop}%
\bibitem [{\citenamefont {Wu}\ and\ \citenamefont {Bruun}(2016)}]{Wu2016}%
  \BibitemOpen
  \bibfield  {author} {\bibinfo {author} {\bibfnamefont {Z.}~\bibnamefont {Wu}}\ and\ \bibinfo {author} {\bibfnamefont {G.~M.}\ \bibnamefont {Bruun}},\ }\href {https://doi.org/10.1103/PhysRevLett.117.245302} {\bibfield  {journal} {\bibinfo  {journal} {Phys. Rev. Lett.}\ }\textbf {\bibinfo {volume} {117}},\ \bibinfo {pages} {245302} (\bibinfo {year} {2016})}\BibitemShut {NoStop}%
\bibitem [{\citenamefont {DeMille}\ \emph {et~al.}(2024)\citenamefont {DeMille}, \citenamefont {Hutzler}, \citenamefont {Rey},\ and\ \citenamefont {Zelevinsky}}]{demille2024quantum}%
  \BibitemOpen
  \bibfield  {author} {\bibinfo {author} {\bibfnamefont {D.}~\bibnamefont {DeMille}}, \bibinfo {author} {\bibfnamefont {N.~R.}\ \bibnamefont {Hutzler}}, \bibinfo {author} {\bibfnamefont {A.~M.}\ \bibnamefont {Rey}},\ and\ \bibinfo {author} {\bibfnamefont {T.}~\bibnamefont {Zelevinsky}},\ }\href {https://doi.org/10.1038/s41567-024-02499-9} {\bibfield  {journal} {\bibinfo  {journal} {Nature Physics}\ }\textbf {\bibinfo {volume} {20}},\ \bibinfo {pages} {741} (\bibinfo {year} {2024})}\BibitemShut {NoStop}%
\bibitem [{\citenamefont {Langen}\ \emph {et~al.}(2024)\citenamefont {Langen}, \citenamefont {Valtolina}, \citenamefont {Wang},\ and\ \citenamefont {Ye}}]{langen2024quantum}%
  \BibitemOpen
  \bibfield  {author} {\bibinfo {author} {\bibfnamefont {T.}~\bibnamefont {Langen}}, \bibinfo {author} {\bibfnamefont {G.}~\bibnamefont {Valtolina}}, \bibinfo {author} {\bibfnamefont {D.}~\bibnamefont {Wang}},\ and\ \bibinfo {author} {\bibfnamefont {J.}~\bibnamefont {Ye}},\ }\href {https://doi.org/10.1038/s41567-024-02423-1} {\bibfield  {journal} {\bibinfo  {journal} {Nature Physics}\ }\textbf {\bibinfo {volume} {20}},\ \bibinfo {pages} {702} (\bibinfo {year} {2024})}\BibitemShut {NoStop}%
\bibitem [{\citenamefont {\ifmmode~\acute{S}\else \'{S}\fi{}mia\l{}kowski}\ and\ \citenamefont {Tomza}(2021)}]{Smialkowski}%
  \BibitemOpen
  \bibfield  {author} {\bibinfo {author} {\bibfnamefont {M.}~\bibnamefont {\ifmmode~\acute{S}\else \'{S}\fi{}mia\l{}kowski}}\ and\ \bibinfo {author} {\bibfnamefont {M.}~\bibnamefont {Tomza}},\ }\href {https://doi.org/10.1103/PhysRevA.103.022802} {\bibfield  {journal} {\bibinfo  {journal} {Phys. Rev. A}\ }\textbf {\bibinfo {volume} {103}},\ \bibinfo {pages} {022802} (\bibinfo {year} {2021})}\BibitemShut {NoStop}%
\bibitem [{\citenamefont {Fleig}\ and\ \citenamefont {DeMille}(2021)}]{fleig2021theoretical}%
  \BibitemOpen
  \bibfield  {author} {\bibinfo {author} {\bibfnamefont {T.}~\bibnamefont {Fleig}}\ and\ \bibinfo {author} {\bibfnamefont {D.}~\bibnamefont {DeMille}},\ }\href {https://iopscience.iop.org/article/10.1088/1367-2630/ac3619} {\bibfield  {journal} {\bibinfo  {journal} {New Journal of Physics}\ }\textbf {\bibinfo {volume} {23}},\ \bibinfo {pages} {113039} (\bibinfo {year} {2021})}\BibitemShut {NoStop}%
\bibitem [{\citenamefont {Sunaga}\ \emph {et~al.}(2019)\citenamefont {Sunaga}, \citenamefont {Abe}, \citenamefont {Hada},\ and\ \citenamefont {Das}}]{sunaga2019merits}%
  \BibitemOpen
  \bibfield  {author} {\bibinfo {author} {\bibfnamefont {A.}~\bibnamefont {Sunaga}}, \bibinfo {author} {\bibfnamefont {M.}~\bibnamefont {Abe}}, \bibinfo {author} {\bibfnamefont {M.}~\bibnamefont {Hada}},\ and\ \bibinfo {author} {\bibfnamefont {B.~P.}\ \bibnamefont {Das}},\ }\href {https://doi.org/10.1103/PhysRevA.99.062506} {\bibfield  {journal} {\bibinfo  {journal} {Phys. Rev. A}\ }\textbf {\bibinfo {volume} {99}},\ \bibinfo {pages} {062506} (\bibinfo {year} {2019})}\BibitemShut {NoStop}%
\bibitem [{\citenamefont {K{\l}os}\ \emph {et~al.}(2022)\citenamefont {K{\l}os}, \citenamefont {Li}, \citenamefont {Tiesinga},\ and\ \citenamefont {Kotochigova}}]{klos2022prospects}%
  \BibitemOpen
  \bibfield  {author} {\bibinfo {author} {\bibfnamefont {J.}~\bibnamefont {K{\l}os}}, \bibinfo {author} {\bibfnamefont {H.}~\bibnamefont {Li}}, \bibinfo {author} {\bibfnamefont {E.}~\bibnamefont {Tiesinga}},\ and\ \bibinfo {author} {\bibfnamefont {S.}~\bibnamefont {Kotochigova}},\ }\href {https://iopscience.iop.org/article/10.1088/1367-2630/ac50ea} {\bibfield  {journal} {\bibinfo  {journal} {New Journal of Physics}\ }\textbf {\bibinfo {volume} {24}},\ \bibinfo {pages} {025005} (\bibinfo {year} {2022})}\BibitemShut {NoStop}%
\bibitem [{\citenamefont {Marc}\ \emph {et~al.}(2023)\citenamefont {Marc}, \citenamefont {Hubert},\ and\ \citenamefont {Fleig}}]{marc2023candidate}%
  \BibitemOpen
  \bibfield  {author} {\bibinfo {author} {\bibfnamefont {A.}~\bibnamefont {Marc}}, \bibinfo {author} {\bibfnamefont {M.}~\bibnamefont {Hubert}},\ and\ \bibinfo {author} {\bibfnamefont {T.}~\bibnamefont {Fleig}},\ }\href {https://doi.org/10.1103/PhysRevA.108.062815} {\bibfield  {journal} {\bibinfo  {journal} {Phys. Rev. A}\ }\textbf {\bibinfo {volume} {108}},\ \bibinfo {pages} {062815} (\bibinfo {year} {2023})}\BibitemShut {NoStop}%
\bibitem [{\citenamefont {Polet}\ \emph {et~al.}(2024)\citenamefont {Polet}, \citenamefont {Chamorro}, \citenamefont {Pa{\v{s}}teka}, \citenamefont {Hoekstra}, \citenamefont {Tomza}, \citenamefont {Borschevsky},\ and\ \citenamefont {Aucar}}]{polet2024p}%
  \BibitemOpen
  \bibfield  {author} {\bibinfo {author} {\bibfnamefont {J.~D.}\ \bibnamefont {Polet}}, \bibinfo {author} {\bibfnamefont {Y.}~\bibnamefont {Chamorro}}, \bibinfo {author} {\bibfnamefont {L.~F.}\ \bibnamefont {Pa{\v{s}}teka}}, \bibinfo {author} {\bibfnamefont {S.}~\bibnamefont {Hoekstra}}, \bibinfo {author} {\bibfnamefont {M.}~\bibnamefont {Tomza}}, \bibinfo {author} {\bibfnamefont {A.}~\bibnamefont {Borschevsky}},\ and\ \bibinfo {author} {\bibfnamefont {I.~A.}\ \bibnamefont {Aucar}},\ }\href {https://doi.org/10.1063/5.0235522} {\bibfield  {journal} {\bibinfo  {journal} {The Journal of Chemical Physics}\ }\textbf {\bibinfo {volume} {161}},\ \bibinfo {pages} {234302} (\bibinfo {year} {2024})}\BibitemShut {NoStop}%
\bibitem [{\citenamefont {Marc}\ and\ \citenamefont {Fleig}(2025)}]{marc2025semi}%
  \BibitemOpen
  \bibfield  {author} {\bibinfo {author} {\bibfnamefont {A.}~\bibnamefont {Marc}}\ and\ \bibinfo {author} {\bibfnamefont {T.}~\bibnamefont {Fleig}},\ }\href {https://doi.org/10.1140/epjd/s10053-025-00967-2} {\bibfield  {journal} {\bibinfo  {journal} {The European Physical Journal D}\ }\textbf {\bibinfo {volume} {79}},\ \bibinfo {pages} {19} (\bibinfo {year} {2025})}\BibitemShut {NoStop}%
\bibitem [{\citenamefont {Uhlenberg}\ \emph {et~al.}(2000)\citenamefont {Uhlenberg}, \citenamefont {Dirscherl},\ and\ \citenamefont {Walther}}]{Uhlenberg2000}%
  \BibitemOpen
  \bibfield  {author} {\bibinfo {author} {\bibfnamefont {G.}~\bibnamefont {Uhlenberg}}, \bibinfo {author} {\bibfnamefont {J.}~\bibnamefont {Dirscherl}},\ and\ \bibinfo {author} {\bibfnamefont {H.}~\bibnamefont {Walther}},\ }\href {https://doi.org/10.1103/PhysRevA.62.063404} {\bibfield  {journal} {\bibinfo  {journal} {Phys. Rev. A}\ }\textbf {\bibinfo {volume} {62}},\ \bibinfo {pages} {063404} (\bibinfo {year} {2000})}\BibitemShut {NoStop}%
\bibitem [{\citenamefont {Ketterle}\ \emph {et~al.}(1993)\citenamefont {Ketterle}, \citenamefont {Davis}, \citenamefont {Joffe}, \citenamefont {Martin},\ and\ \citenamefont {Pritchard}}]{Ketterle1993}%
  \BibitemOpen
  \bibfield  {author} {\bibinfo {author} {\bibfnamefont {W.}~\bibnamefont {Ketterle}}, \bibinfo {author} {\bibfnamefont {K.~B.}\ \bibnamefont {Davis}}, \bibinfo {author} {\bibfnamefont {M.~A.}\ \bibnamefont {Joffe}}, \bibinfo {author} {\bibfnamefont {A.}~\bibnamefont {Martin}},\ and\ \bibinfo {author} {\bibfnamefont {D.~E.}\ \bibnamefont {Pritchard}},\ }\href {https://doi.org/10.1103/PhysRevLett.70.2253} {\bibfield  {journal} {\bibinfo  {journal} {Phys. Rev. Lett.}\ }\textbf {\bibinfo {volume} {70}},\ \bibinfo {pages} {2253} (\bibinfo {year} {1993})}\BibitemShut {NoStop}%
\bibitem [{\citenamefont {Falke}\ \emph {et~al.}(2006)\citenamefont {Falke}, \citenamefont {Tiemann}, \citenamefont {Lisdat}, \citenamefont {Schnatz},\ and\ \citenamefont {Grosche}}]{PhysRevA.74.032503}%
  \BibitemOpen
  \bibfield  {author} {\bibinfo {author} {\bibfnamefont {S.}~\bibnamefont {Falke}}, \bibinfo {author} {\bibfnamefont {E.}~\bibnamefont {Tiemann}}, \bibinfo {author} {\bibfnamefont {C.}~\bibnamefont {Lisdat}}, \bibinfo {author} {\bibfnamefont {H.}~\bibnamefont {Schnatz}},\ and\ \bibinfo {author} {\bibfnamefont {G.}~\bibnamefont {Grosche}},\ }\href {https://doi.org/10.1103/PhysRevA.74.032503} {\bibfield  {journal} {\bibinfo  {journal} {Phys. Rev. A}\ }\textbf {\bibinfo {volume} {74}},\ \bibinfo {pages} {032503} (\bibinfo {year} {2006})}\BibitemShut {NoStop}%
\bibitem [{\citenamefont {Nath}\ \emph {et~al.}(2013)\citenamefont {Nath}, \citenamefont {Easwaran}, \citenamefont {Rajalakshmi},\ and\ \citenamefont {Unnikrishnan}}]{nath2013}%
  \BibitemOpen
  \bibfield  {author} {\bibinfo {author} {\bibfnamefont {D.}~\bibnamefont {Nath}}, \bibinfo {author} {\bibfnamefont {R.~K.}\ \bibnamefont {Easwaran}}, \bibinfo {author} {\bibfnamefont {G.}~\bibnamefont {Rajalakshmi}},\ and\ \bibinfo {author} {\bibfnamefont {C.~S.}\ \bibnamefont {Unnikrishnan}},\ }\href {https://doi.org/10.1103/PhysRevA.88.053407} {\bibfield  {journal} {\bibinfo  {journal} {Phys. Rev. A}\ }\textbf {\bibinfo {volume} {88}},\ \bibinfo {pages} {053407} (\bibinfo {year} {2013})}\BibitemShut {NoStop}%
\bibitem [{\citenamefont {Salomon}\ \emph {et~al.}(2014)\citenamefont {Salomon}, \citenamefont {Fouch{\'e}}, \citenamefont {Wang}, \citenamefont {Aspect}, \citenamefont {Bouyer},\ and\ \citenamefont {Bourdel}}]{salomon2014gray}%
  \BibitemOpen
  \bibfield  {author} {\bibinfo {author} {\bibfnamefont {G.}~\bibnamefont {Salomon}}, \bibinfo {author} {\bibfnamefont {L.}~\bibnamefont {Fouch{\'e}}}, \bibinfo {author} {\bibfnamefont {P.}~\bibnamefont {Wang}}, \bibinfo {author} {\bibfnamefont {A.}~\bibnamefont {Aspect}}, \bibinfo {author} {\bibfnamefont {P.}~\bibnamefont {Bouyer}},\ and\ \bibinfo {author} {\bibfnamefont {T.}~\bibnamefont {Bourdel}},\ }\href {https://iopscience.iop.org/article/10.1209/0295-5075/104/63002/meta} {\bibfield  {journal} {\bibinfo  {journal} {Europhysics letters}\ }\textbf {\bibinfo {volume} {104}},\ \bibinfo {pages} {63002} (\bibinfo {year} {2014})}\BibitemShut {NoStop}%
\bibitem [{\citenamefont {Badr}\ \emph {et~al.}(2004)\citenamefont {Badr}, \citenamefont {Plimmer}, \citenamefont {Juncar}, \citenamefont {Himbert}, \citenamefont {Silver},\ and\ \citenamefont {Rovera}}]{badr2004continuous}%
  \BibitemOpen
  \bibfield  {author} {\bibinfo {author} {\bibfnamefont {T.}~\bibnamefont {Badr}}, \bibinfo {author} {\bibfnamefont {M.}~\bibnamefont {Plimmer}}, \bibinfo {author} {\bibfnamefont {P.}~\bibnamefont {Juncar}}, \bibinfo {author} {\bibfnamefont {M.}~\bibnamefont {Himbert}}, \bibinfo {author} {\bibfnamefont {J.}~\bibnamefont {Silver}},\ and\ \bibinfo {author} {\bibfnamefont {G.}~\bibnamefont {Rovera}},\ }\href {https://link.springer.com/article/10.1140/epjd/e2004-00118-y} {\bibfield  {journal} {\bibinfo  {journal} {The European Physical Journal D-Atomic, Molecular, Optical and Plasma Physics}\ }\textbf {\bibinfo {volume} {31}},\ \bibinfo {pages} {3} (\bibinfo {year} {2004})}\BibitemShut {NoStop}%
\bibitem [{\citenamefont {Dzuba}\ \emph {et~al.}(2021)\citenamefont {Dzuba}, \citenamefont {Allehabi}, \citenamefont {Flambaum}, \citenamefont {Li},\ and\ \citenamefont {Schiller}}]{Dzuba2021}%
  \BibitemOpen
  \bibfield  {author} {\bibinfo {author} {\bibfnamefont {V.~A.}\ \bibnamefont {Dzuba}}, \bibinfo {author} {\bibfnamefont {S.~O.}\ \bibnamefont {Allehabi}}, \bibinfo {author} {\bibfnamefont {V.~V.}\ \bibnamefont {Flambaum}}, \bibinfo {author} {\bibfnamefont {J.}~\bibnamefont {Li}},\ and\ \bibinfo {author} {\bibfnamefont {S.}~\bibnamefont {Schiller}},\ }\href {https://doi.org/10.1103/PhysRevA.103.022822} {\bibfield  {journal} {\bibinfo  {journal} {Phys. Rev. A}\ }\textbf {\bibinfo {volume} {103}},\ \bibinfo {pages} {022822} (\bibinfo {year} {2021})}\BibitemShut {NoStop}%
\bibitem [{\citenamefont {Porsev}\ \emph {et~al.}(2025)\citenamefont {Porsev}, \citenamefont {Filin}, \citenamefont {Cheung},\ and\ \citenamefont {Safronova}}]{Porsev2025}%
  \BibitemOpen
  \bibfield  {author} {\bibinfo {author} {\bibfnamefont {S.~G.}\ \bibnamefont {Porsev}}, \bibinfo {author} {\bibfnamefont {D.}~\bibnamefont {Filin}}, \bibinfo {author} {\bibfnamefont {C.}~\bibnamefont {Cheung}},\ and\ \bibinfo {author} {\bibfnamefont {M.~S.}\ \bibnamefont {Safronova}},\ }\href {https://doi.org/10.1103/PhysRevA.111.062809} {\bibfield  {journal} {\bibinfo  {journal} {Phys. Rev. A}\ }\textbf {\bibinfo {volume} {111}},\ \bibinfo {pages} {062809} (\bibinfo {year} {2025})}\BibitemShut {NoStop}%
\end{thebibliography}%

\end{document}